WE economy: Potential of mutual aid distribution based on moral responsibility and risk vulnerability

Short title: WE economy: mutual aid distribution

Takeshi Kato[1*]

[1] Hitachi Kyoto University Laboratory, Open Innovation Institute, Kyoto University, Kyoto, Japan

* Corresponding author

E-mail: kato.takeshi.3u@kyoto-u.ac.jp (TK)

# Abstract

Reducing wealth inequality and disparity is a global challenge. To accomplish this, the economic systems that produce inequality must be transformed. The economic system is mainly divided into (1) gift and reciprocity, (2) power and redistribution, (3) market exchange, and (4) mutual aid without reciprocal obligations. The current inequality stems from a capitalist economy consisting of (2) and (3). To sublimate (1), which is the human economy, to (4), the concept of a "mixbiotic society" has been proposed in the philosophical realm. This is a society in which free and diverse individuals, "I," mix with each other, recognize their respective "fundamental incapability" and sublimate them into "WE" solidarity. The economy in this society must have moral responsibility as a coadventurer and consideration for vulnerability to risk. Therefore, I focus on two factors of mind perception: moral responsibility and risk vulnerability, and propose a novel model of wealth distribution between two agents following an econophysical approach. Specifically, I developed a joint-venture model in which profits and losses are distributed based on their own factors; a redistribution model in which wealth stocks are redistributed periodically based on their own factors in the joint-venture model; and a "WE economy" model in which profits and losses are distributed based on the ratio of each other's factors. A simulation comparison of a combination of the joint ventures and redistribution with the WE economies reveals that WE economies are effective in reducing inequality and resilient in normalizing wealth distribution as advantages, and susceptible to free riders as disadvantages. However, this disadvantage can be compensated for by fostering consensus and fellowship, and by complementing it with joint ventures. Although this study does not reflect the real economy, because it is a basic model, it essentially presents the effectiveness of moral responsibility, the complementarity between the WE economy and the joint economy, and the direction of the economy toward reducing inequality. Future challenges are to develop the WE economy model based on real economic analysis and economic psychology, as well as to promote WE economy fieldwork for worker coops and platform cooperatives to realize a desirable mixbiotic society.

## Introduction

Wealth inequality and disparity have become major social issues around the world. According to the World Inequality Report 2022, the top 1% richest people account for 38% of the world's wealth [1], and according to the World Economic Forum, only 8 men of the top have the same wealth as the poorest 3.6 billion people [2]. The global Gini index has reached 0.7 [1], well above the warning level 0.4 for social unrest [3]. Solving the inequality problem is an urgent issue because social unrest creates a vicious cycle that lowers productivity, increases inequality, and further fuels social unrest [4].

The United Nations Sustainable Development Goals include, as Goal 10, reducing inequality, promoting social, economic, and political inclusion, and fiscal and social policies that promote equality, and as Goals 1, 2, 8, and 16, eradicating poverty, zero hunger, inclusive economic growth, and fair and inclusive institutions, respectively [5]. At the World Economic Forum, a gathering of political and business leaders, the reduction of inequality is also on the main agenda [6]. To achieve these goals, it is essential to formulate policies and institutions based on the economic relationships that produce inequality—that is, the mode of wealth exchange.

The economist Polanyi has identified three modes of economic relations: (1) reciprocity, (2) redistribution, and (3) market exchange [7]. Anthropologist Graeber presents (2') hierarchy, (3') exchange, and (4') foundational communism as three moral principles involved in economic relations [8], and philosopher Karatani presents four modes of exchange: (1") reciprocity, (2") plunder and redistribution, (3") commodity exchange, and (4") advanced recovery of reciprocity [9]. Of these, (1)(1') is a human gift economy with an obligation of return, (2)(2')(2") is a power economy with tax collection and redistribution, (3)(3')(3") is a market economy with non-human exchange of goods and money, and (4')(4") is a human mutual-aid economy without an obligation of return and sublimated from a gift economy. The capitalist economy that produces the current inequality is a combination of a power economy and a market economy [8, 9], and Graeber and Karatani advocate a transformation to a (4')(4") human economy as a prescription for the inequality problem. Note that the Islamic economy is a combination of redistribution based on the morals of the Islamic code instead of power (*waqf*, *sadaqah*, and *zakat*) and a joint economy that prohibits interest (*mudaraba*, *murabaha*, and *salam*) [10, 11]. It is a capitalist economic alternative and a possible stepping stone to a human economy [8].

Philosopher Deguchi introduces the concept of a "mixbiotic society," a further development of the symbiotic society, as a social vision corresponding to (4')(4") above [12]. This is a society in which free and diverse individuals, "I," mix with each other, recognize their respective "fundamental incapability" and sublimate them into "WE" solidarity [12, 13]. In a mixbiotic society, individuals entrust and cooperate with each other in a "WE" community. Regarding the human economy in a mixbiotic society—I call it the "WE economy"—Deguchi states in his book [14]:

- Good WE: Fellowship, equality, hollowness without power center, cooperativity, voluntary participation, softened WE.

- Bad WE: Totalitarianism, exclusivism toward the outside, peer pressure toward the inside, hardened WE.

- Good WE are coadventurers who participate while accepting the risk together.

- WE members qualify as coadventurers (risk takers) in that they have vulnerability and frangibility.

- Coadventurers in the same boat may have economic class divisions, but they are on equal footing as a community of destiny.

- Coadventurers are weighted according to their moral responsibility, but the weighting is only a quantitative differentiation of the received profit and the associated risk (latch).

- The risk-return allocation is increased or decreased but not monopolized, and everyone receives a return for the amount of risk taken.

To summarize Deguchi's discourse, two things are important in the "WE Economy": moral responsibility and risk vulnerability. Here, the dimensions of mind perception are informative. According to psychologists Gray and Wegner, mind perception can be divided into two main dimensions: agency (capacities such as self-control, morality, and memory) and experience (capacities such as hunger, fear, and pain) [15]. Then, as shown in S1 Fig in the Supporting information, characters such as a fetus, a baby, a girl, an adult man, an adult woman, a man in persistent vegetative state, and a dead woman are mapped on a two-dimensional plane. Agency corresponds to moral agent (giver of moral action: moral responsibility) and experience to moral patient (recipient of pain: risk vulnerability) [16, 17]. Based on these, the WE economy make the coadventurers undertake risk (latch) in accordance with their moral responsibility and allocates returns in consideration of its responsibility and risk vulnerability.

What potential does the WE economy have for reducing wealth inequality? In examining this, I use an econophysical approach. According to reviews in this field, various wealth distribution models have been proposed based on the analogy of kinetic energy exchange of gaseous particles and other dynamics, and wealth distributions and inequalities such as exponential, power, gamma, and delta distributions have been considered [18, 19]. Recent examples of studies have examined income and inheritance taxes [20], social class and inheritance [21], tax exemptions for the poor [22], contributions of surplus stock [23], interest business, joint ventures and redistribution [24], and redistribution and mutual aid [25]. Among these models, to compare with the WE economy as coadventurers, it seems appropriate to refer to the joint-venture model, which is the most similar to this in the way wealth is distributed.

Therefore, in this study, I formulate a new mathematical model of the WE economy based on moral responsibility and risk vulnerability referring to the above joint-venture model [24], and simulate wealth distribution and inequality. Then, by comparing joint ventures and WE economies, I aim to gain insights to lead the capitalist economy, which generates inequality, to the WE economy. The remainder of this paper is organized as follows: In the Methods section, I first cite the conventional joint-venture model and then present a redistribution model based on moral responsibility and risk vulnerability in the joint-venture model and a distribution model based on responsibility and vulnerability in the WE economy. The Results section presents simulation results for the wealth distribution and the Gini index of inequality. The Discussion section discusses, based on the results, the real-world applicability of the joint ventures with redistribution and the WE economies as alternatives to capitalist economies. Finally, future challenges include the research issues and empirical fieldworks.

# Methods

## Moral responsibility and risk vulnerability

Firstly, using the agency and experience scores by Gray and Wegner shown in S1 Fig [15] in the Supporting information, I establish values for moral responsibility and risk vulnerability. Excluding a fetus, a man in persistent vegetative state, and a dead woman, who are not involved in economic activities, a baby, a girl, an adult man and an adult woman roughly ride on a straight line on a two-dimensional plane. From baby to adult, agency, which corresponds to moral responsibility, roughly changes from $0.2$ to $0.8$, and experience, which corresponds to risk vulnerability, roughly changes from $1$ to $0.8$. The age-specific population distribution includes a stationary type with a constant population of each age group, an expansive type with a large population of young people (population growth), and a constrictive type with a small population of young people (population decline) [26]. Here I assume the stationary type, not the expansive type with short life expectancy or the constrictive type with extreme aging. That is, assuming an even distribution of the number of people from babies to adults, the moral responsibility $\rho_{Mi}$ and risk vulnerability $\rho_{Mi}$ of the $i$-th agent ($i = 1, 2, \cdots, N$) among $N$ agents can be expressed as shown in Eq (1). Note that $\rho_i$ is a multiplication of $\rho_{Mi}$ and $\rho_{Ri}$.

$$\rho_{Mi} = 0.2 + \frac{0.8}{N} \cdot i,$$

$$\rho_{Ri} = 1 - \frac{0.2}{N} \cdot i, \tag{1}$$

$$\rho_i = \rho_{Mi} \cdot \rho_{Ri}.$$

## Joint-venture and redistribution models

First, as an econophysical model, I refer to the basic joint-venture model presented in the literature [24] (called the JV-B model). In the JV-B model, two agents $i$ and $j$ ($i \neq j, i, j = 1, 2, \cdots, N$) are randomly selected at time $t$ among $N$ agents. Both agents have wealth $m_i(t)$ and $m_j(t)$, respectively, and a common savings rate $\lambda$. Both agents contribute their wealth, excluding savings, to the joint venture, and wealth is distributed according to the wealth $(1 - \lambda) \cdot m_i(t)$ and $(1 - \lambda) \cdot m_j(t)$ contributed by each agent and the profit/loss ratio $\delta$. The wealth $m_i(t + 1)$ and $m_j(t + 1)$ of the two agents $i$ and $j$ at time $t + 1$ are expressed as shown in Eq (2), respectively.

$$m_i(t + 1) = \lambda \cdot m_i(t) + (1 + \delta) \cdot (1 - \lambda) \cdot m_i(t);$$
$$m_j(t + 1) = \lambda \cdot m_j(t) + (1 + \delta) \cdot (1 - \lambda) \cdot m_j(t). \tag{2}$$

In the JV-B model, all wealth excluding savings was contributed to the joint venture. But in the joint-venture model in this study, I model the two agents $i$ and $j$ contributing wealth according to their moral responsibilities $\rho_{Mi}$ and $\rho_{Mj}$ (called the JV-M model). That is, two agents $i$ and $j$ contribute wealth $(1 - \lambda) \cdot \rho_{Mi} \cdot m_i(t)$ and $(1 - \lambda) \cdot \rho_{Mj} \cdot m_j(t)$, respectively, and wealth is distributed according to the profit/loss rate $\delta$. The wealth $m_i(t + 1)$ and $m_j(t + 1)$ at time $t + 1$ are expressed as shown in Eq (3), respectively.

$$m_i(t+1) = \lambda \cdot m_i(t) + (1-\lambda) \cdot (1-\rho_{Mi}) \cdot m_i(t)$$

$$+(1+\delta) \cdot (1-\lambda) \cdot \rho_{Mi} \cdot m_i(t);$$

$$m_j(t+1) = \lambda \cdot m_j(t) + (1-\lambda) \cdot (1-\rho_{Mj}) \cdot m_j(t) \qquad (3)$$

$$+(1+\delta) \cdot (1-\lambda) \cdot \rho_{Mj} \cdot m_j(t).$$

Next, I formulate a redistribution model for the JV-M model (called JV-M-M model) by referring to the redistribution model in the literature [24]. In the JV-M-M model, each of the $N$ agents contributes wealth $\xi \cdot \rho_{Mi} \cdot m_i(t)$ according to its transfer ratio $\xi$ and moral responsibility $\rho_{Mi}$ every redistribution period $t_p$, and the wealth $\sum_{k=1}^{N} \rho_{Mk} \cdot m_k(t)$ collected from the $N$ agents is redistributed to each agent according to its moral responsibility ratio $\rho_{Mi}/\sum_k \rho_{Mk}$. The wealth $m_i(t+\Delta)$ of agent $i$ at time $t+\Delta$ after redistribution is expressed as shown in Eq (4).

$$m_i(t+\Delta) = (1 - \xi \cdot \rho_{Mi}) \cdot m_i(t) + \xi \cdot \frac{\rho_{Mi}}{\sum_k \rho_{Mk}} \cdot \sum_{k=1}^{N} \rho_{Mk} \cdot m_k(t). \qquad (4)$$

In the JV-M-M model, redistribution was made according to moral responsibility, but for comparison, I model redistribution according to agent $i$'s risk vulnerability $\rho_{Ri}$ (called the JV-M-R model). In the JV-M-R model, the wealth collected from $N$ agents is redistributed to each of them according to their risk vulnerability ratio $\rho_{Ri}/\sum_k \rho_{Rk}$. The wealth $m_i(t+\Delta)$ of agent $i$ at time $t+\Delta$ after redistribution is expressed as shown in Eq (5).

$$m_i(t+\Delta) = (1 - \xi \cdot \rho_{Mi}) \cdot m_i(t) + \xi \cdot \frac{\rho_{Ri}}{\sum_k \rho_{Rk}} \cdot \sum_{k=1}^{N} \rho_{Mk} \cdot m_k(t). \qquad (5)$$

Similar to the JV-M-M and JV-M-R models, I model the redistribution according to both moral responsibility $\rho_{Mi}$ and risk vulnerability $\rho_{Ri}$ of agent $i$ (called the JV-M-MR model). In the JV-M-MR model, the wealth collected from $N$ agents is redistributed to each according to the ratio $\rho_i / \sum_k \rho_k$ using $\rho_i = \rho_{Mi} \cdot \rho_{Ri}$ in Eq (1). The wealth $m_i(t + \Delta)$ of agent $i$ at time $t + \Delta$ after redistribution is expressed as shown in Eq (6).

$$m_i(t + \Delta) = (1 - \xi \cdot \rho_{Mi}) \cdot m_i(t) + \xi \cdot \frac{\rho_i}{\sum_k \rho_k} \cdot \sum_{k=1}^{N} \rho_{Mk} \cdot m_k(t). \qquad (6)$$

## WE economy models

In the JV-M model, wealth is distributed according to the wealth $(1 - \lambda) \cdot \rho_{Mi} \cdot m_i(t)$ and $(1 - \lambda) \cdot \rho_{Mj} \cdot m_j(t)$ contributed by two agents $i$ and $j$, respectively. In the WE economy, to distribute as coadventurers or community of destiny, the wealth contributed by the two agents $i$ and $j$ according to their moral responsibilities $\rho_{Mi}$ and $\rho_{Mj}$ is collected once as $(1 - \lambda) \cdot \left( \rho_{Mi} \cdot m_i(t) + \rho_{Mj} \cdot m_j(t) \right)$ and distributed according to their respective moral responsibility ratio $\rho_{Mi} / (\rho_{Mi} + \rho_{Mj})$ and $\rho_{Mj} / (\rho_{Mi} + \rho_{Mj})$ (called WE-M-M model). The wealth $m_i(t + 1)$ and $m_j(t + 1)$ of the two agents $i$ and $j$ at time $t + 1$ are expressed as shown in Eq (7), respectively.

$$
\begin{aligned}
m_i(t + 1) &= \lambda \cdot m_i(t) + (1 - \lambda) \cdot (1 - \rho_{Mi}) \cdot m_i(t) \\
&\quad + (1 + \delta) \cdot (1 - \lambda) \cdot \frac{\rho_{Mi}}{\rho_{Mi} + \rho_{Mj}} \cdot \left( \rho_{Mi} \cdot m_i(t) + \rho_{Mj} \cdot m_j(t) \right); \\
m_j(t + 1) &= \lambda \cdot m_j(t) + (1 - \lambda) \cdot \left( 1 - \rho_{Mj} \right) \cdot m_j(t). \\
&\quad + (1 + \delta) \cdot (1 - \lambda) \cdot \frac{\rho_{Mj}}{\rho_{Mi} + \rho_{Mj}} \cdot \left( \rho_{Mi} \cdot m_i(t) + \rho_{Mj} \cdot m_j(t) \right).
\end{aligned}
\qquad (7)
$$

In the WE-M-M model, redistribution was made according to moral responsibility, but for comparison, I model redistribution according to agent $i$'s risk vulnerability $\rho_{Mi}$ (called the WE-M-R model). In the WE-M-R model, the wealth contributed by the two agents $i$ and $j$ according to their moral responsibilities $\rho_{Mi}$ and $\rho_{Mj}$ is distributed according to their risk vulnerability ratios $\rho_{Ri}/(\rho_{Ri} + \rho_{Rj})$ and $\rho_{Rj}/(\rho_{Ri} + \rho_{Rj})$. The wealth $m_i(t+1)$ and $m_j(t+1)$ at time $t+1$ are expressed as shown in Eq (8), respectively.

$$
\begin{aligned}
m_i(t+1) &= \lambda \cdot m_i(t) + (1-\lambda) \cdot (1-\rho_{Mi}) \cdot m_i(t) \\
&+ (1+\delta) \cdot (1-\lambda) \cdot \frac{\rho_{Ri}}{\rho_{Ri}+\rho_{Rj}} \cdot \left(\rho_{Mi} \cdot m_i(t) + \rho_{Mj} \cdot m_j(t)\right); \\
m_j(t+1) &= \lambda \cdot m_j(t) + (1-\lambda) \cdot (1-\rho_{Mj}) \cdot m_j(t) \\
&+ (1+\delta) \cdot (1-\lambda) \cdot \frac{\rho_{Rj}}{\rho_{Ri}+\rho_{Rj}} \cdot \left(\rho_{Mi} \cdot m_i(t) + \rho_{Mj} \cdot m_j(t)\right).
\end{aligned} \tag{8}
$$

Similar to the WE-M-M and WE-M-R models, I model distribution according to both the moral responsibilities $\rho_{Mi}, \rho_{Mj}$ and risk vulnerabilities $\rho_{Ri}, \rho_{Rj}$ of the two agents $i$ and $j$ (called the WE-M-MR model). In the WE-M-MR model, using $\rho_i = \rho_{Mi} \cdot \rho_{Ri}$ in Eq (1), the wealth contributed by the two agents $i$ and $j$ according to their moral responsibilities $\rho_{Mi}$ and $\rho_{Mj}$ is distributed according to their ratios $\rho_i/(\rho_i + \rho_j)$ and $\rho_j/(\rho_i + \rho_j)$. The wealth $m_i(t+1)$ and $m_j(t+1)$ at time $t+1$ are expressed as shown in Eq (9), respectively.

$$
\begin{aligned}
m_i(t+1) &= \lambda \cdot m_i(t) + (1-\lambda) \cdot (1-\rho_{Mi}) \cdot m_i(t) \\
&+ (1+\delta) \cdot (1-\lambda) \cdot \frac{\rho_i}{\rho_i+\rho_j} \cdot \left(\rho_{Mi} \cdot m_i(t) + \rho_{Mj} \cdot m_j(t)\right); \\
m_j(t+1) &= \lambda \cdot m_j(t) + (1-\lambda) \cdot (1-\rho_{Mj}) \cdot m_j(t) \\
&+ (1+\delta) \cdot (1-\lambda) \cdot \frac{\rho_j}{\rho_i+\rho_j} \cdot \left(\rho_{Mi} \cdot m_i(t) + \rho_{Mj} \cdot m_j(t)\right).
\end{aligned} \tag{9}
$$

## Impact of free riders

I now refer to the JV-M model in Eq (3) and the WE-M-M model in Eq (7) to examine the impact of free riders who are not cooperative in joint ventures and WE economies. Assuming that one agent $j$ of the two agents $i$ and $j$ contributes wealth only by multiplying its moral responsibility $\rho_{Mj}$ by the ratio $r_f$, Eqs (3) and (7) can be rewritten as Eqs (10) and (11), respectively. For convenience, I call to the model combining Eq (10) with the redistribution in Eq (4) as the JV-M-M-FR model and the model in Eq (11) as the WE-M-M-FR model. Note that the redistribution in the JV-M-M-FR model does not take into account the impact of free riders, because the redistribution is considered to be institutionally done for everyone.

$$m_i(t+1) = \lambda \cdot m_i(t) + (1-\lambda) \cdot (1-\rho_{Mi}) \cdot m_i(t) + (1+\delta) \cdot$$
$$(1-\lambda) \cdot \rho_{Mi} \cdot m_i(t);$$
$$m_j(t+1) = \lambda \cdot m_j(t) + (1-\lambda) \cdot \left(1 - r_f \cdot \rho_{Mj}\right) \cdot m_j(t)$$
$$+ (1+\delta) \cdot (1-\lambda) \cdot r_f \cdot \rho_{Mj} \cdot m_j(t). \tag{10}$$

$$m_i(t+1) = \lambda \cdot m_i(t) + (1-\lambda) \cdot (1-\rho_{Mi}) \cdot m_i(t)$$
$$+ (1+\delta) \cdot (1-\lambda) \cdot \frac{\rho_{Mi}}{\rho_{Mi}+\rho_{Mj}} \cdot \left(\rho_{Mi} \cdot m_i(t) + r_f \cdot \rho_{Mj} \cdot m_j(t)\right);$$
$$m_j(t+1) = \lambda \cdot m_j(t) + (1-\lambda) \cdot \left(1 - r_f \cdot \rho_{Mj}\right) \cdot m_j(t)$$
$$+ (1+\delta) \cdot (1-\lambda) \cdot \frac{\rho_{Mj}}{\rho_{Mi}+\rho_{Mj}} \cdot \left(\rho_{Mi} \cdot m_i(t) + r_f \cdot \rho_{Mj} \cdot m_j(t)\right). \tag{11}$$

## Calculation of Gini Index

The Gini index is a well-known parameter for assessing wealth inequality [27] and is calculated by drawing a Lorenz curve and an equal distribution line [28]. There are various inequality indices calculated from Lorenz curves [29], but here I use the most common Gini index. By the operation $Sort(m_i(t))$ in Eq (12), the wealth $m_i(t)$ $(i = 1, 2, \cdots, N)$ of the $N$ agents at time $t$ is sorted from smallest to largest, and the $k$-th wealth from smallest is set as $\mu_k(t)$, and the Gini index $g$ is calculated.

$$\mu_k(t) \in Sort\big(m_i(t)\big),$$

$$g = \frac{2 \cdot \sum_{k=1}^{N} k \cdot \mu_k(t)}{N \cdot \sum_{k=1}^{N} \mu_k(t)} - \frac{N+1}{N}. \tag{12}$$

The Gini index $g = 0$ if the wealth of the $N$ agents is equally distributed, and $g = 1$ if it is delta-distributed (all wealth is concentrated in only one agent). In other words, the more unequal the distribution is, the larger the Gini index is.

## Results

To begin, common parameters for the simulations of the joint-venture model, the redistribution model, and the WE economy model are set. The number of agents is $N = 1,000$ (which does not affect the relative calculation of wealth distribution or Gini index), and time is run from $t = 0$ to $10^6$ in increments of $1$. The initial distribution of the wealth of the $N$ agents at time $t = 0$ is equally $m_i(0) = 1$ ($i = 1, 2, \cdots, N$) (this will be discussed later). The savings rate is $\lambda = 0.25$, referring to the world's gross savings (as a percentage of GDP) [30]. With respect to the profit/loss ratio $\delta$, while the average return of the stock index is about 8%, there are large fluctuations exceeding ±10% [31], compared to an average return of only about 2% for investors [32]. In other words, to account for the fact that business profits and losses fluctuate both positively and negatively, here I set a uniform random number in the range $-0.1 \leq \delta \leq 0.1$ for every time $t$ for the profit/loss rate $\delta$. For the redistribution period $t_p$ and transfer rate $\xi$ of the joint-venture model, I use the combination $t_p = 10^4$ and $\xi = 0.5$, where the Gini index $g$ is relatively small, referring to the literature [24].

Figs 1A and 1B show the calculation results for the JV-B model of conventional joint ventures expressed in Eq (1), Figs 1C and 1D show the JV-M model of joint ventures based on moral responsibility expressed in Eq (2), and Figs 1E and 1F show the WE-M-M model of the WE economy based on moral responsibility expressed in Eq (7). Figs 1A, 1C, and 1E show the frequency distribution at time $t = 10^4$, $10^5$, and $10^6$ with wealth $m$ on the horizontal axis and frequencies on the vertical axis. Figs 1B, 1D, and 1F plot the wealth $m$ of each agent at time $t = 10^6$ with agent number $\#$ on the horizontal axis.

In the JV-B and JV-M models, as time $t$ increases, the frequency distributions of the wealth in Figs 1A and 1C approach the $m = 0$ side, while the wealth $m$ of the agents in Figs 1B and 1D are widely distributed. Literature [24] shows that the joint-venture model without redistribution gradually approaches a delta distribution (Gini index $g = 1$). In Fig 1D, there are no plots near $m = 0$ for $\#$ roughly in the range from 1 to 300. This is because in the JV-M model, the smaller $\#$ is, the smaller the moral responsibility is, and thus the wealth contribution is suppressed and less subject to fluctuations in profits and losses. Compared to the JV-B and JV-M models, the WE-M-M model in Figs 1E and 1F concentrates the distribution of wealth in the neighborhood of $m = 1$. This is because, as can be seen by comparing Eqs (3) and (7), in the WE-M-M model the wealth contributed by the two agents was once added together and then distributed according to the ratio of moral responsibility.

Fig 2 shows the calculation results for the combinations of joint-venture models and redistribution models. Figs 2A and 2B show the JV-M-M model combining Eqs (3) and (4), Figs 2C and 2D show the JV-M-R model combining Eqs (3) and (5), and Figs 2E and 2F show the JV-M-MR model combining Eqs (3) and (6). Compared to the JV-M models in Figs 1C and 1D, in Figs 2A, 2B, 2E, and 2F, the distribution range of wealth $m$ is narrower due to redistribution and is concentrated near $m = 1$. In Figs 2B and 2F, the variance of wealth $m$ is larger for larger # because larger wealth contributions are subject to fluctuations in profits and losses. In contrast to Fig 2B, in Fig 2F, wealth $m$ on the side with smaller # is slightly larger than on the side with larger # because of redistribution based on both moral responsibility and risk vulnerability. In Fig 2C the frequency distribution is skewed toward $m = 0$ and in Fig 2D the wealth $m$ is smaller for larger #. This is because, as can be seen from Eqs (1) and (5), the wealth contributed according to moral responsibility was redistributed according to risk vulnerability, resulting in an imbalance between the contribution and redistribution of wealth.

Fig 3 shows the calculation results for the WE economy models. Figs 3A and 3B show the WE-M-M model of Eq (7) (the same as Figs 1E and 1F, respectively), Figs 3C and 3D show the WE-M-R model of Eq (8), and Figs 3E and 3F show the WE-M-MR model of Eq (9). Compared to the joint-venture and redistribution models of Figs 2A, 2B, 2E, and 2F, the WE economy models of Figs 3A, 3B, 3E, and 3F further concentrate the distribution of wealth $m$ near $m = 1$. In other words, WE economies have the advantage of reducing inequality with respect to joint ventures. This is due to the wealth-adding effect in WE economies, as described in Fig 1. The skewed distribution of wealth $m$ in Figs 3C and 3D is due to an imbalance between the contribution and distribution of wealth, as described in Figs 2C and 2D. The difference between Figs 3B and 3F is due to the inclusion of risk vulnerability in the distribution, similar to the difference between Figs 2B and 2F. Note that Figs 2B and 3B, Figs 2D and 3D, and Figs 2E and 3E have nearly equivalent trends, suggesting that the joint-venture redistribution and the WE economy distribution have similar inequality reduction effects.

Fig 4 shows the calculation results when the initial distribution of wealth $m_i(0)$ ($i = 1, 2, \cdots, N$) is changed from an even distribution with all 1's to a real uniform random number distribution between 0 and 2. Figs 4A and 4B show the JV-M-M-IR model with a changed initial distribution of the JV-M-M model, and Figs 4C and 4D show the WE-M-M-IR model with a changed initial distribution of the WE-M-M model. The JV-M-M-IR model in Figs 4A and 4B for the JV-M-M model in Figs 2A and 2B, and the WE-M-M-IR model in Figs 4C and 4D for the WE-M-M model in Figs 3A and 3B show almost equivalent results, respectively. These indicates that the redistribution of joint ventures and the WE economies have the resilience to converge the distribution of wealth $m$ over time $t$. Moreover, the resilience of the WE Economies is higher than that of the redistribution of joint ventures.

Fig 5 shows the calculation results when free riders are considered. The ratio of free riders in Eqs (10) and (11) is assumed to be $r_f = 0.5$ simply for calculation. Figs 5A and 5B show the JV-M-M-FR model combining Eq (10) and the redistribution in Eq (4), while Figs 5C and 5D show the WE-M-M-FR model in Eq (11). Comparing the JV-M-M model in Figs 2A and 2B with the JV-M-M-FR model in Figs 5A and 5B, there is almost no difference in the distribution of wealth $m$ in both models. This is because, as can be seen in Eq (10), in the joint venture model, the effect of a free-rider agent only spills over to that agent itself. In contrast, the WE-M-M-FR model in Figs 5C and 5D has a wider distribution of wealth $m$ than the WE-M-M model in Figs 3A and 3B. This is because, as can be seen in Eq (11), in the WE economy model, the reduction in the free-rider agent's wealth contribution affects both of the two agents.

This indicates that WE economies have the disadvantage of being more susceptible to free riders than joint ventures.

Fig 6 shows the results of the Gini index calculation. Fig 6 shows the change of the Gini index $g$ with time $t$, where the horizontal axis is time $t$ and the vertical axis (logarithm) is the Gini index $g$. In Fig 6A, the Gini index $g$ of the JV-M model without redistribution (gray line) tends toward $1$ with time $t$. The Gini indices $g$ for the joint-venture model with redistribution and the WE economy model converge to a constant value less than the social unrest warning level 0.4 [3]. The WE-M-M model (green) is the smallest, followed by the WE-M-MR model (dark green), which includes risk vulnerability. The JV-M-M (blue) and JV-M-MR (dark blue) models, and the JV-M-R (purple) and WE-M-R (blue-green) models have nearly equivalent values, respectively. The overall trend is that the WE economy model has a smaller Gini index than the joint venture model, and the redistribution of joint ventures and the WE economies, including risk vulnerability, has a slightly larger Gini index.

In Fig 6B, the JV-M-M (blue) and JV-M-M-IR (purple), and the WE-M-M (green) and WE-M-M-IR (blue-green) models, which have different initial distributions of wealth $m$, converge to approximately the same value of Gini index $g$ over time $t$, respectively. With respect to the impact of free riders, the Gini index $g$ is slightly smaller in the JV-M-M-FR model (dark blue) than in the JV-M-M model (blue). This is due to the reduced contribution of free-rider wealth, which reduces the impact of profits and losses. This has been shown in the literature [23, 25] as a proportional relationship between the amount of contributions and the Gini index (i.e., the Gini index decreases as the amount of contributions is reduced). The Gini index $g$ of the WE-M-M-FR model (dark green) is larger than that of the WE-M-M model (green). This has already been explained as the reason for the difference between Figs 5C and 5D relative to Figs 3A and 3B.

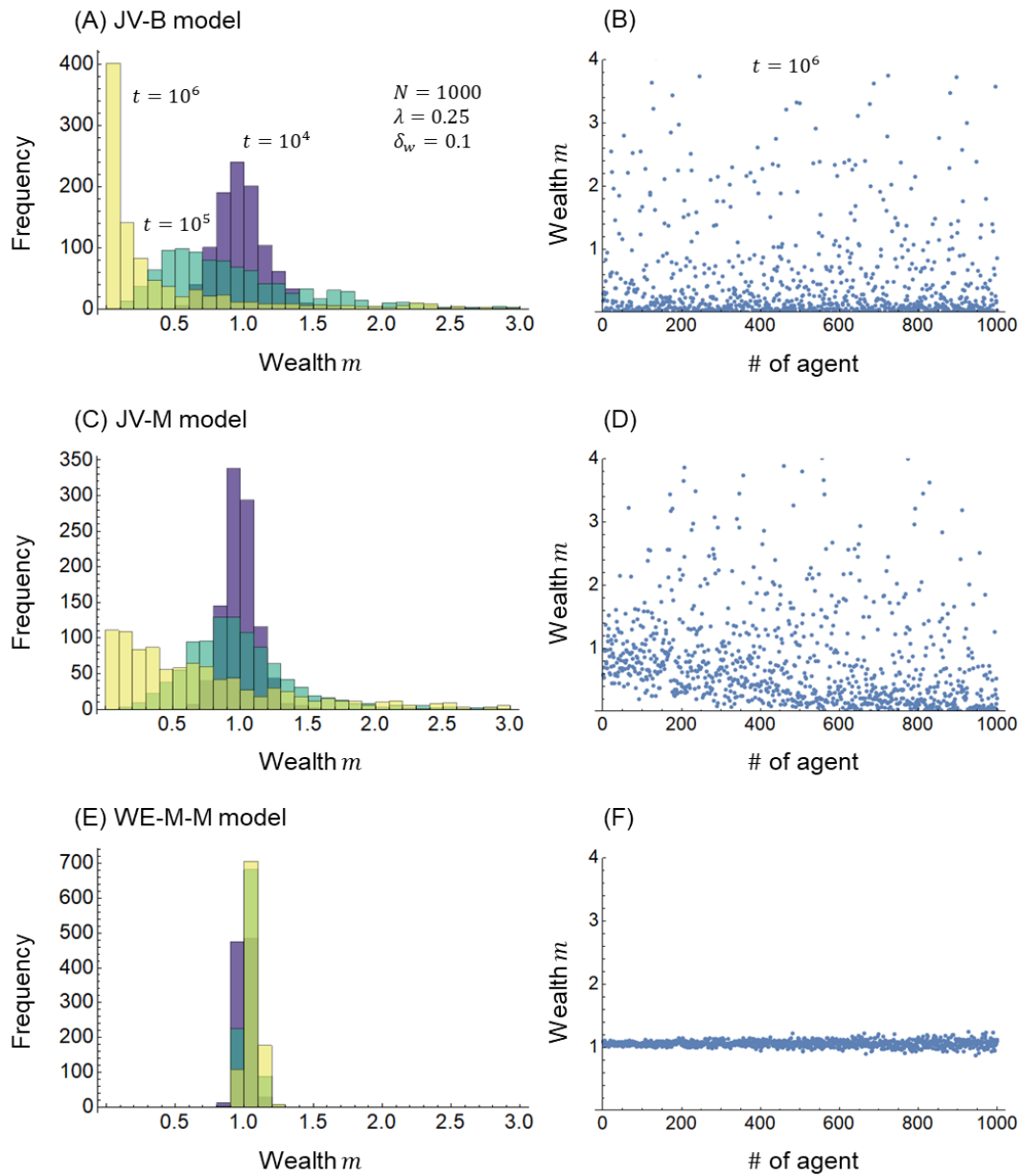

**Fig 1. Wealth distribution.**

(A)(B) JV-B model, (C)(D) JV-M model, and (E)(F) WE-M-M model.

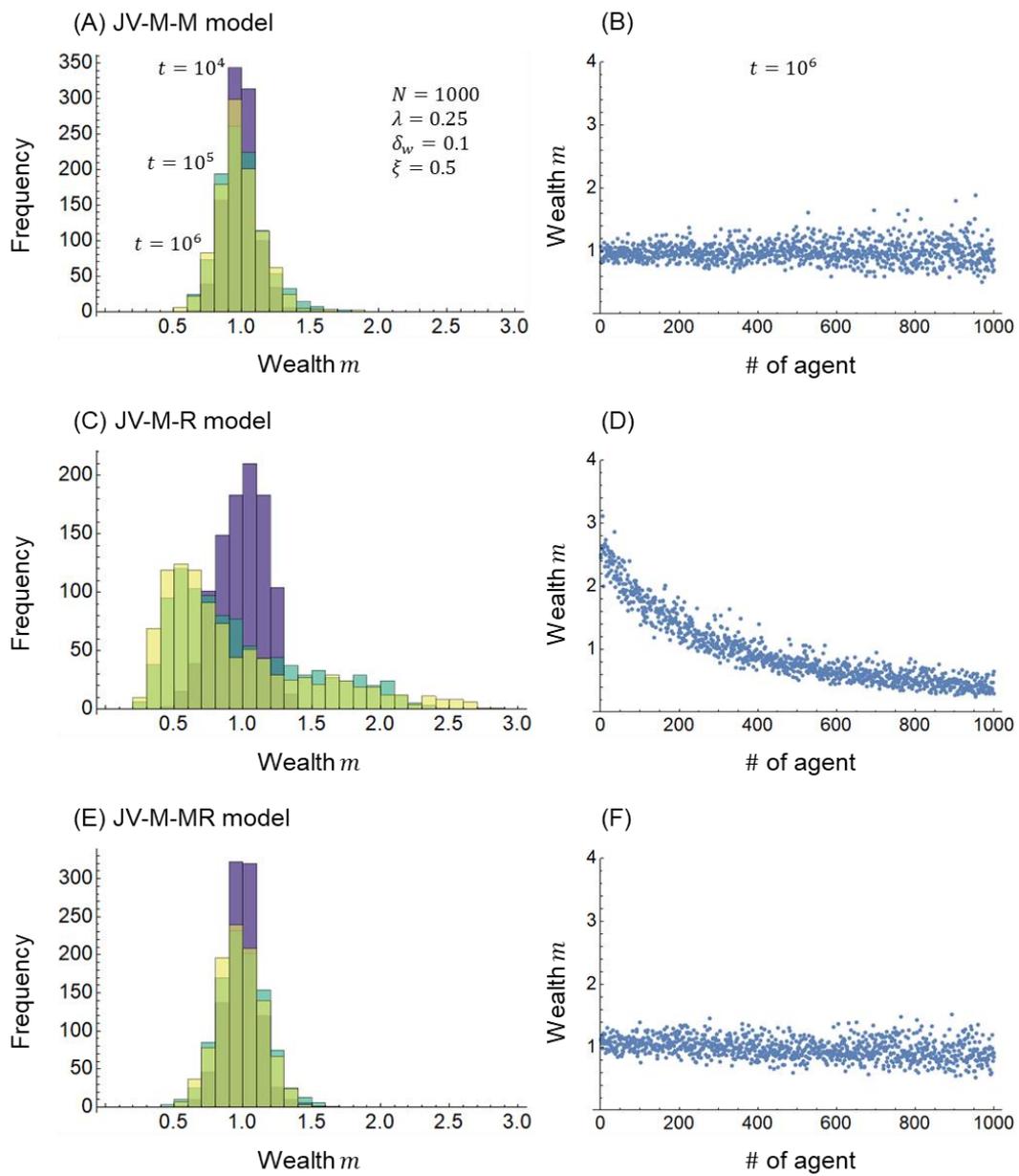

**Fig 2. Wealth distribution.**

(A)(B) JV-M-M model, (C)(D) JV-M-R model, and (E)(F) JV-M-MR model.

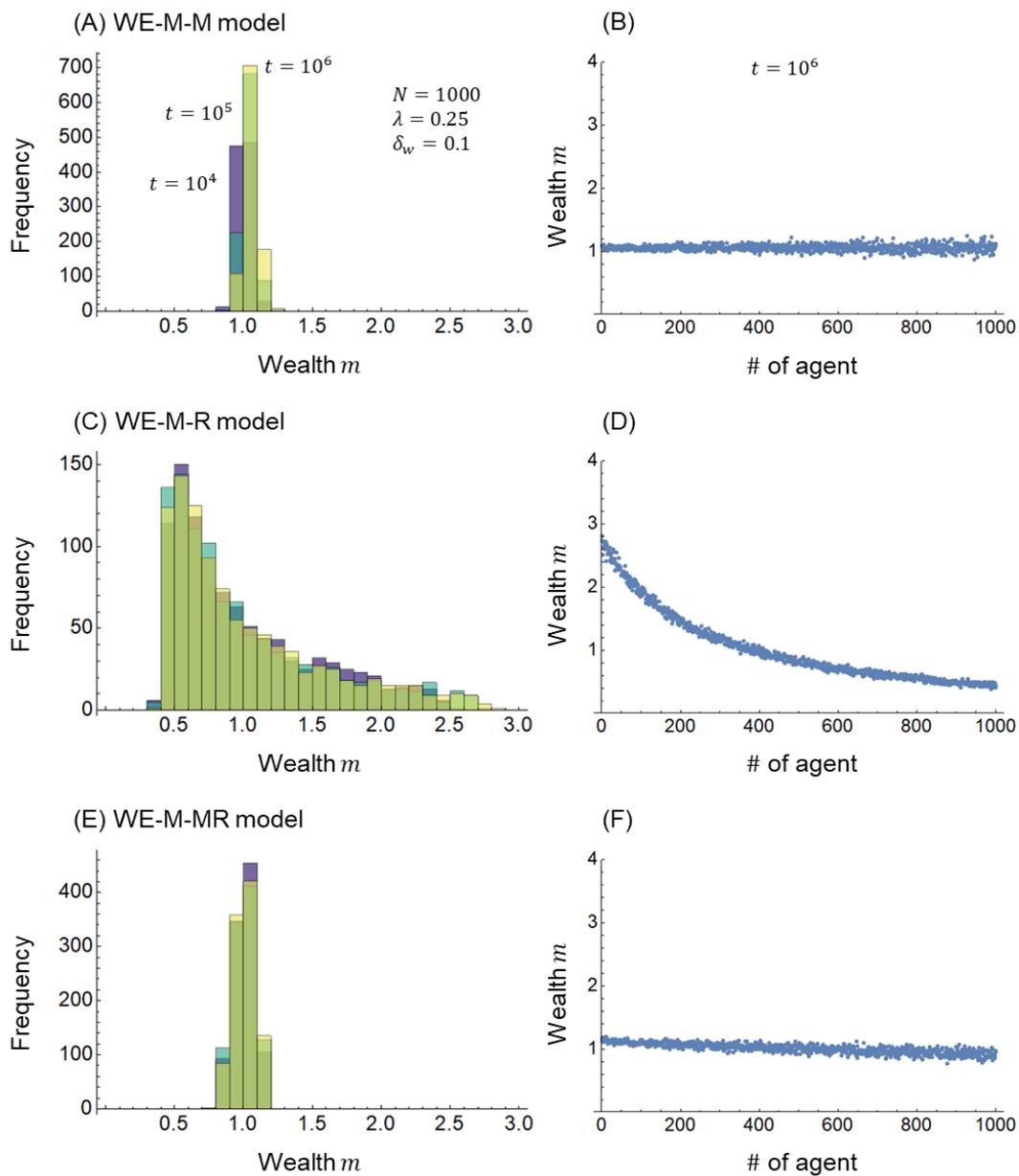

**Fig 3. Wealth distribution.**

(A)(B) WE-M-M model, (C)(D) WE-M-R, and (E)(F) WE-M-MR model.

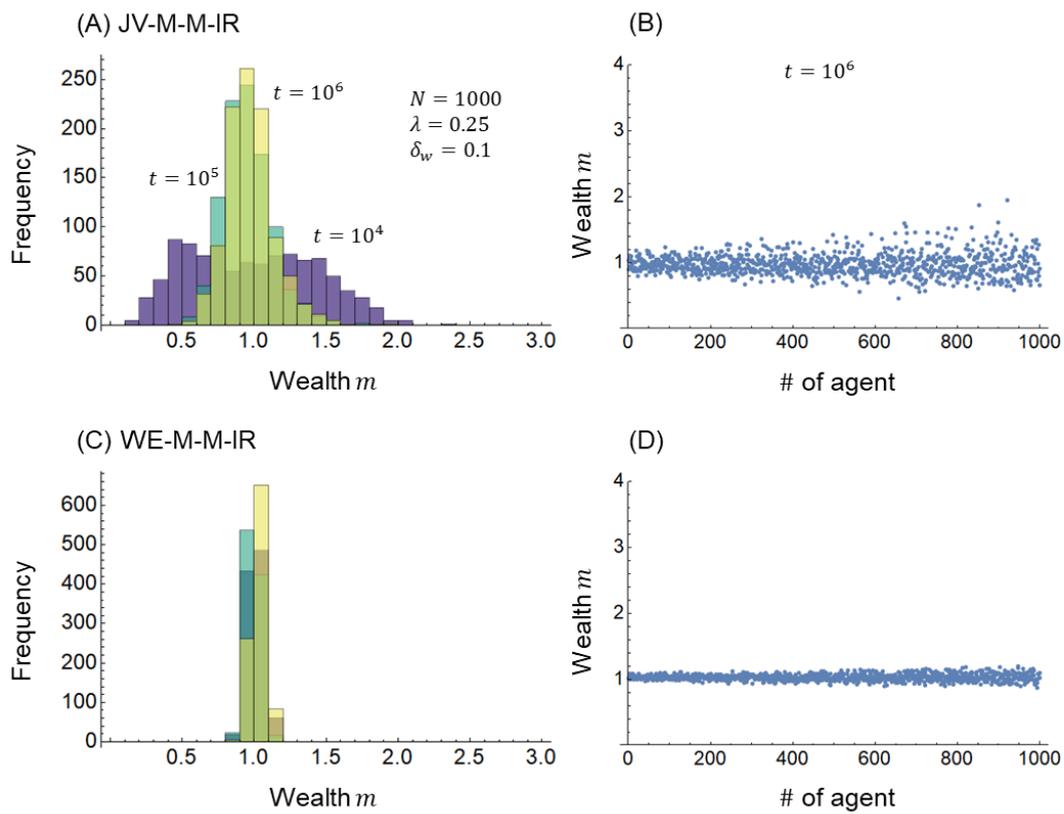

**Fig 4. Wealth distribution.**

(A)(B) JV-M-M-IR model and (C)(D) WE-M-M-IR model.

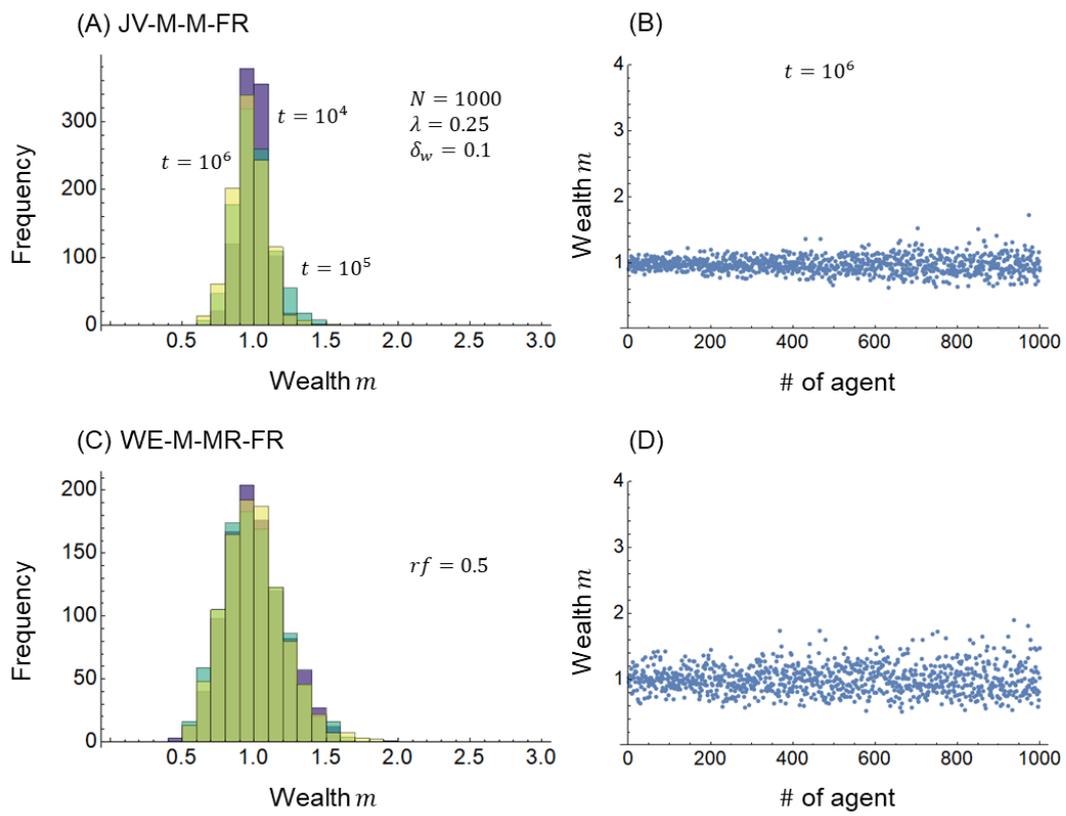

**Fig 5. Wealth distribution.**

(A)(B) JV-M-M-FR model and (C)(D) WE-M-M-FR model.

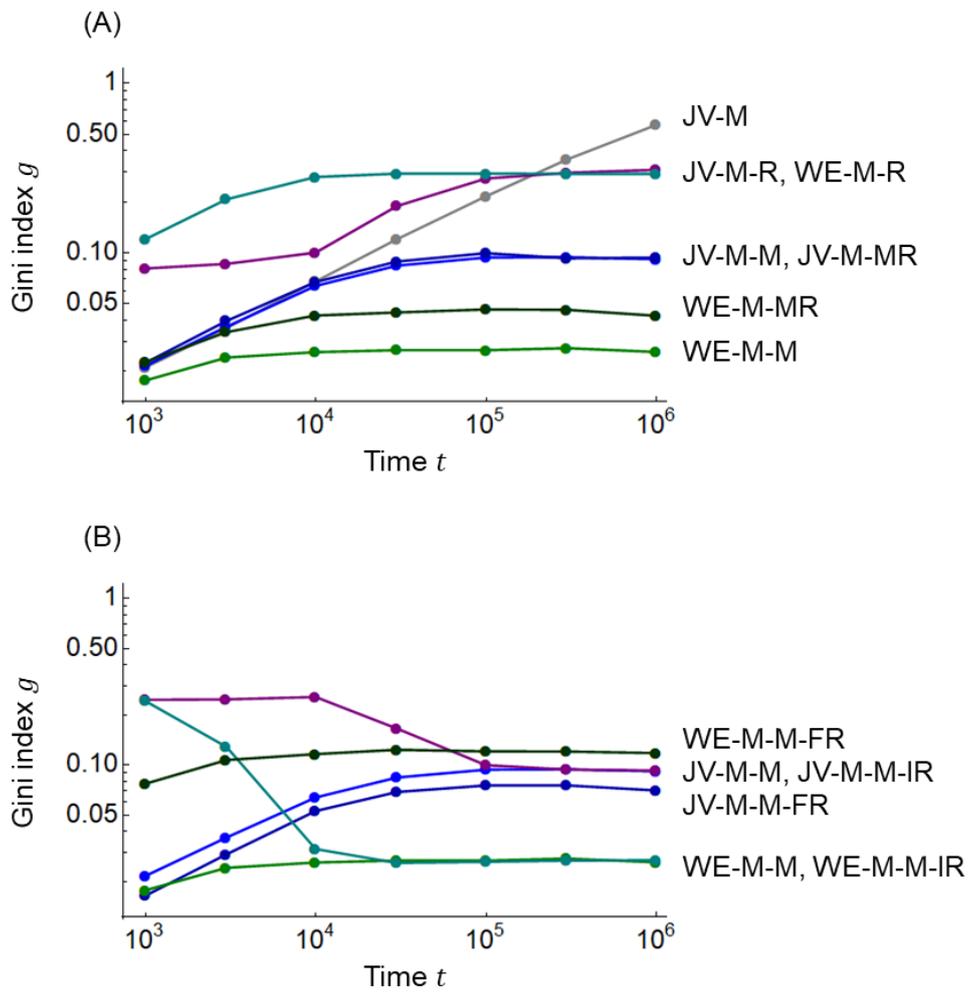

**Fig 6. Gini index.**

(A) JV-M; JV-M-M, JV-M-R, JV-M-MR; WE-M-M, WE-M-R, and WE-M-MR models. (B) JV-M-M, JV-M-M-IR, JV-M-M-FR; WE-M-M, WE-M-M-IR, and WE-M-M-FR models.

# Discussion

The calculation results for WE economies and joint ventures show that moral responsibility can be an economic instrument instead of inhuman money, and the comparison between the two newly shows that WE economies have a higher potential to reduce wealth inequality than joint ventures. This is because in a joint venture, the profits and losses from the wealth contributed by each person are distributed according to their own responsibilities, whereas in a WE economy, the profits and losses from the wealth contributed as a coadventurer or community of destiny are distributed according to each other's moral responsibility ratio. This study is the first to demonstrate the effectiveness of not only the mixbiotic society shown by the philosopher Deguchi [12–14], but also the foundational communism shown by Graeber [8] and the advanced recovery of reciprocity shown by Karatani [9], inherited from Mauss's theory of gifts [33] and Kropotkin's theory of mutual aid [34], by an econophysical approach.

The WE Economy model is similar to a hunter-gatherer society, by analogy. That is, in this model, each tribal member brings food obtained through hunting and gathering to the communal square once, and then distributes the food according to moral responsibility and risk vulnerability. In contrast, the joint-venture model is more akin to an agrarian society. In this model, villagers cooperate with each other in water management and seedling planting, but food from the land belongs to the villager who owns the land, and the storage of food creates inequality. The redistribution model for joint ventures is similar to the role of the rich and the chief in providing relief to the poor based on moral responsibility. Note that these are a reminder of the combination of joint economy *(mudaraba, murabaha,* and *salam)* and redistribution *(waqf, sadaqah,* and *zakat*) in the Islamic economy.

The effectiveness of wealth distribution and redistribution based on moral responsibility in both WE economies and joint ventures suggests that a shift from a money economy to a credit economy is possible. The money economy ia based on equivalent exchange divorced from the social context, and the credit economy is based on moral and mutual aid. Taking a long-term view of human history, Graeber notes that money and credit economies have alternated, that in the Middle Ages the transition from a money economy to a credit economy occurred from Islamic societies, and that we are currently in a transition from a money economy to a credit economy [8]. The results of this work will support Graeber's vision and encourage a transformation from the capitalist economy, which currently produces inequality, to a humanistic mutual-aid economy.

The resilience of the WE economy and joint ventures/redistribution based on moral responsibility to the distribution of wealth shows that it has the potential to break the vicious cycle of social unrest, as described at the beginning of the Introduction section, and stabilize society. Furthermore, a distribution of wealth based on both moral responsibility and risk vulnerability tilts toward people who are more vulnerable. This will contribute to a more inclusive society by, for example, helping children taking the future, young people leading the next generation, and the socially vulnerable.

However, WE economies have the disadvantage of being more susceptible to free riders than joint ventures with redistribution. This can be viewed, in other words, as a governance issue in a coadventurer or community of destiny. Economist Ostrom cites collective choice of operating rules, effective monitoring, and graduated sanctions as design principles for the commons [35]. Of these, there is a fear that monitoring and sanctions will lead to the bad WE (totalitarianism, peer pressure) indicated by Deguchi, but collective choice of operating rules, including free riders, or consensus building to recognize each other's moral responsibility, will be important.

Economist Bowles, in his quite book "The Moral Economy," presents the trilemma of Pareto efficiency, preference neutrality, and voluntary participation, and cites limiting preference neutrality as a solution to this trilemma [36]. The limitation of preference neutrality is, for example, the moral sensitization of free riders and the fostering of fellowship and cooperation. Once this is done, both respect for voluntary participation and the efficiency of the WE economy will be ensured, as indicated by Deguchi. Moreover, since societies and communities are multi-layered, an apparent free-rider in one community can move to another based on the voluntary participation.

Given the disadvantages of WE economies, even if they are compensated for by collective choice and restrictions on preference neutrality, WE economies would be suitable for relatively small-scale local societies and communities with common moral and social norms. This is because WE economies require mutual recognition of moral responsibility and risk vulnerability. Specific examples include worker cooperatives [37], where workers hold labor, investment and management, and platform cooperatives [38, 39], which are based on joint ownership and democratic decision-making by users and workers. As an implementation of the WE economy through information technology, initiatives such as the Social Co-Operating System [40], which combines an operational loop that promotes cooperative behavior with a collegial loop that supports consensus building, would be useful.

In societies larger than WE economies, joint ventures with redistribution based on moral responsibility would be complementary and effective. In other words, it would constitute a multi-layered network of WE economies and joint ventures/redistribution, and perform the joint ventures/redistribution between the WE economies. However, the redistribution should be based on norms and morals, as in the Islamic economy, rather than on hierarchies as shown by Graeber [8] or plunder by power as shown by Karatani [9]. The Islamic economy is highly compatible with the WE economy in that it emphasizes a real, face-to-face, and mutual-aid economy and balances self-interest and altruism through various institutions [10, 11]. Just as Islamic societies triggered a shift to a credit economy in the Middle Ages [8], so too in modern times the Islamic economy could be a stepping stone to a shift to a capitalist economic alternative.

Note that the simulations were conducted using moral responsibility and risk vulnerability riding on a straight line on a two-dimensional plane based on the literature [15], but the basic relationship and trends between WE economies and joint ventures/redistribution should not change even if both are distributed on the plane. It is known that the mind perception changes with mental states [41, 42] and with non-human objects [43, 44]. In the future, it is expected that artificially intelligent agents with morality will emerge [14]. Although the mind perception is expected to change depending on the context of economic actors, economic activities, and social relations, the fundamental importance of the WE economy should remain the same.

This study only models the basic WE economy and the joint ventures/redistribution, and the absolute values of the parameters and calculation results do not necessarily reflect the real economy. However, because it is a basic model that discards details, it essentially presents the effectiveness of moral responsibility, the complementarity of the WE economy and the joint economy, and the direction of economic transformation toward reducing wealth inequality. As future challenges, there remain analytical studies based on modeling and parameterization that reflect the real economy, psychological research on the mind perception in economic activities, empirical research through fieldwork on economic activities based on moral responsibility, and social movements to spread the WE economy and transform a money economy into a credit economy.

# Acknowledgements

This research was conducted as part of the "Toward Better 'Smart WE': From East Asian Humanities and Social Sciences to a Value Multi-Layered Society" project (JSPS Topic-Setting Program to Advance Cutting-Edge Humanities and Social Sciences Research Grant Number JPJS00122679495). I would like to thank Professor Yasuo Deguchi of Kyoto University, the principal investigator of this project, whose ideas of "fundamental incapability," "Self as WE," and "mixbiotic society" directly motivated this study. Furthermore, I would also like to thank my colleagues at the Hitachi Kyoto University Laboratory of the Kyoto University Open Innovation Institute for their ongoing cooperation.

# Supporting information

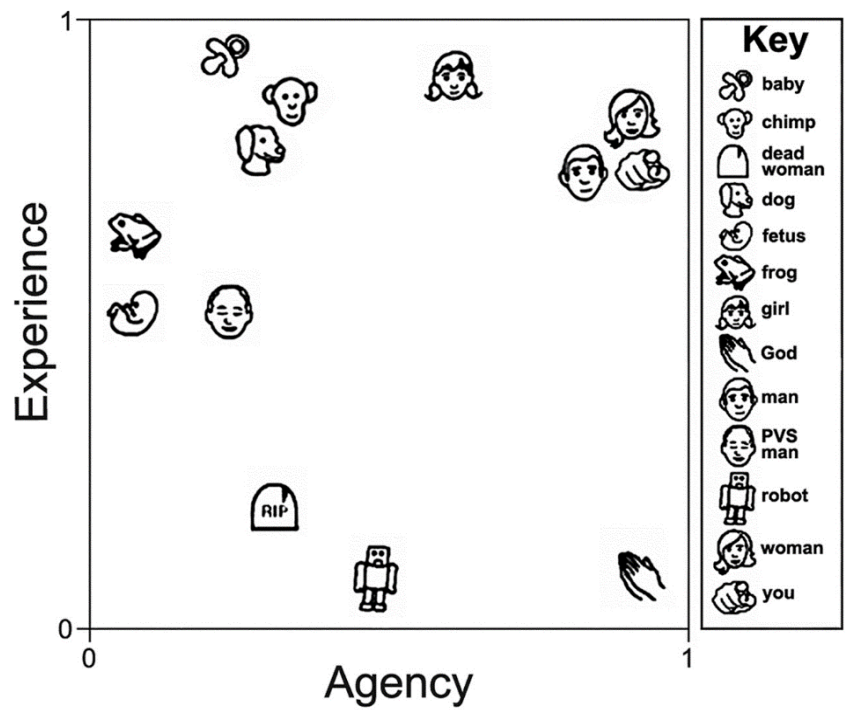

**S1 Fig. Character factor scores on two dimensions of mind perception.**

(Gray, et al. 2007 [15]) Copyright Clearance Center's RightsLink® License Number:

5683990802298